\documentclass[useAMS,usenatbib]{mn2e}

\usepackage{graphicx}
\usepackage{amsmath}
\usepackage{amssymb}
\usepackage{hyperref}
\usepackage{breakurl}

\usepackage{lineno}
\def\m15{M15}

\title[Deep observations of \m15 with MAGIC]
{Deep observations of the globular cluster \m15 with the MAGIC telescopes}

\author[MAGIC Collaboration]{
MAGIC Collaboration:
V.~A.~Acciari$^{1}$, 
S.~Ansoldi$^{2,20}$, 
L.~A.~Antonelli$^{3}$, 
A.~Arbet Engels$^{4}$, \newauthor
D.~Baack$^{5}$,
A.~Babi\'c$^{6}$,
B.~Banerjee$^{7}$,
U.~Barres de Almeida$^{8}$,
J.~A.~Barrio$^{9}$, \newauthor
J.~Becerra Gonz\'alez$^{1}$
% ...
\\
(Affiliations can be found after the references)
}
\begin{document}

\date{Accepted . Received ; in original form }

\pagerange{\pageref{firstpage}--\pageref{lastpage}} \pubyear{2015}

\maketitle
\clearpage
\label{firstpage}
\begin{abstract}
A population of globular clusters (GCs) has been recently established by the {\it Fermi}-LAT telescope as a new class of GeV $\gamma$-ray sources. 
Leptons accelerated to TeV energies, in the inner magnetospheres of MSPs or in their wind regions, should produce $\gamma$-rays through the inverse Compton scattering in the dense radiation field from the huge population of stars.
We have conducted deep observations of the globular cluster \m15 with the MAGIC telescopes and used 165\,hrs in order to search for $\gamma$-ray emission.
A strong upper limit on the TeV $\gamma$-ray flux $<3.2\times 10^{-13}\mathrm{cm^{-2}s^{-1}}$ above 300\,GeV ($<0.26\%$ of the Crab nebula flux) has been obtained. 
We interpret this limit as a constraint on the efficiency of the acceleration of leptons in the magnetospheres of the MSPs.
We constrain the injection rate of relativistic leptons, $\eta_{\rm e}$, from the MSPs magnetospheres and their surrounding. 
We conclude that $\eta_{\rm e}$ must be  
lower than expected from the modelling of high energy processes in MSP inner magnetospheres.
For leptons accelerated with the power law spectrum in the MSP wind regions, $\eta_{\rm e}$ is constrained to be much lower than derived for the wind regions around classical pulsars. 
These constraints are valid for the expected range of magnetic field strengths within the GC and for the range of likely energies of leptons injected from the inner magnetospheres, provided that the leptons are not removed from the globular cluster very efficiently due to advection process. 
We discuss consequences of  
these constraints for the models of radiation processes around millisecond pulsars.
\end{abstract}

\section{Introduction}

About 160 globular clusters (GCs) are gathered in a spherical halo around the center of the Galaxy within a radius of $\sim$10 kpc \citep{har96}. GCs contain a huge number of old and low mass stars, with a total mass up to $\sim$10$^6$~M$_\odot$ in a volume with the radius of a few parsecs, and also evolved, compact objects such as millisecond pulsars (MSPs), cataclysmic variables, and low-mass X-ray binaries. 
Twenty GCs have been recently identified with GeV $\gamma$-ray sources discovered by the {\it Fermi}-LAT telescope \citep{ab09a,ab10, kon10,tam11,zha16}. In the case of two GCs, M28 and NGC~6624, exceptionally energetic MSPs have also been discovered: B1821-24 and J1823-3021A, respectively.
The GeV $\gamma$-ray emission has been observed from these two GCs, modulated with the periods of the pulsars within the cluster \citep{fre11, joh13}. 
Moreover, the spectra observed from GCs have very similar features to those observed from isolated MSPs, i.e. the spectra are flatter than -2 with a cut-off at a few GeV \citep{ab10}. These observations support
that the GeV $\gamma$-ray emission from GCs is very likely produced due to a cumulative emission of the MSP population, 
as proposed by \citep{ha05,ve08,ve09}.  
The basic features of these observations and their consequences for the MSP population in GCs are reviewed by e.g. \cite{bed11} and \cite{tam16}. 

MSPs are expected to inject energetic leptons from their inner magnetospheres in the form of pulsar winds as observed in the case of the pulsars formed in the core collapse supernovae (so called "classical pulsars"). 
These leptons can be injected directly with TeV energies. Alternatively, they may reach such energies as a result of re-acceleration in the pulsar wind regions, in collisions of those winds among themselves or with the winds of the GC stars. Similarities with phenomena around classical pulsars and MSPs have triggered the attention of the telescopes
sensitive in the TeV $\gamma$-ray energies. However, in most cases only upper limits on the TeV $\gamma$-ray flux have been reported, e.g. from Omega Centauri \citep{ka07}, 47~Tuc \citep{ah09}, M13 \citep{an09}, and  \m15, M13, and M5 \citep{mc09}. The upper limits of the TeV flux for several individual GCs (and also stacked upper limits) have also been reported for the case of point-like or extended sources, mostly for the GCs not detected by {\it Fermi}-LAT in GeV $\gamma$-rays \citep{abr13}. Only in the case of GC Ter~5, longer observations ($\sim$ 90 hrs) with the H.E.S.S. telescopes resulted in the discovery of an extended TeV $\gamma$-ray source in the direction of this GC \citep{abr11}. Surprisingly, the centre of the TeV source is shifted from the centre of Ter~5 by the distance comparable to the dimension of the GC. In fact, such asymmetry might be observed in the case of the non-spherical propagation
of TeV leptons around Ter~5 \citep{bs14}.

Detailed MSP models for the TeV $\gamma$-ray emission from GCs have been developed already before the above-mentioned ${\it Fermi}$-LAT discovery of GeV $\gamma$-ray emission from GCs. Two general models for the injection of energetic leptons were considered, either mono-energetic injection directly from the MSPs (e.g. \citealp{bs07, ve09, che10, zbr13, bss16}) or with a power-law distribution as a result of re-acceleration in the pulsar wind collision regions (e.g. \citealp{bs07, kop13, bss16, ndi18}).
These works assumed that TeV $\gamma$-rays originate in the Inverse Compton (IC) scattering process of low energy radiation (optical from the GC, Cosmic Microwave Background (CMB), or the infra-red and optical radiation from the Galactic disk) by leptons accelerated by MSPs. 
The most recent developments of numerical codes computing the $\gamma$-ray emission from GCs take into account effects related to different diffusion scenarios for 
energetic particles \citep{ndi18}, the advection of leptons from GCs due to Red Giant (RG) winds and 
non-homogeneous injection of leptons into the GCs \citep{bss16}. Modelling results, if confronted with observations, should allow us to constrain the processes occurring in the MSP magnetospheres, acceleration of leptons within GCs, and their transport within and around GCs.  
Note that other phenomena, such as supernova remnants accelerating hadrons \citep{dom11} or electrons accelerated in magnetized White Dwarfs \citep{bed12}, might also contribute to the high energy emission from GCs.

The IC model also predicts synchrotron emission from the same population of leptons which might be observable under favourable conditions between the radio and soft X-rays.
In fact, in the {\it Chandra} observations of the GC Ter~5, the existence of an extended, non-thermal X-ray source centred on the GC core has been reported \citep{edc10,cl11}. A similar result has been reported in the case of 47~Tuc \citep{wu14}. Earlier observations have also reported evidence of X-ray emission from some GCs, extended over a few arcmin, which has been interpreted as a result of the interaction of the wind from the GC with the surrounding medium \citep{har82, ok07}. However, such X-ray sources have not been detected in the direction of a few other GCs \citep{ed12}.
Other models for the TeV $\gamma$-ray emission from GCs have been also proposed (see e.g. \citealp{che10, dom11, bed12}).

In summary, the production of TeV $\gamma$-rays in GCs seems to be unavoidable. However, the expected level of emission depends on several parameters, such as injection rate and spectra of leptons from the MSP magnetospheres or the propagation of energetic particles in the complex medium. These parameters are at present not well constrained. Their limitation will have important consequences for observational plans with the future generation of telescopes such as the Cherenkov Telescope Array (CTA, \citealp{acha13}).   
Therefore, we performed extensive observations of the GC \m15 with the MAGIC telescopes. 
In Section~\ref{sec:m15}, we introduce the basic parameters of \m15. 
In Section~\ref{sec:magicm15}, we present the MAGIC observations of \m15.
In Section~\ref{sec:model}, we confront the results of the MAGIC observations with the model of VHE $\gamma$-ray emission of \m15. 
We conclude the paper with final remarks in Section~\ref{sec:conc}. 
\section{Globular cluster \m15}\label{sec:m15}

\m15 belongs to the class of core collapsed, luminous GCs.
Its total stellar luminosity is $7\times 10^5$ L$_\odot$, the core radius 0.43\,pc, the half mass radius 3.04\,pc, and the distance from Earth of 10.4\,kpc \citep{har96}. Eight MSPs have been discovered up to now within \m15 \citep{fre15}.

We observed \m15 with the MAGIC telescopes since it is the only GC discovered at GeV $\gamma$-ray energies by {\it Fermi}-LAT \citep{zha16}, which can be observed at low zenith angles ($Zd_{\rm min}\sim 16^\circ35'$) from La Palma.
Its $\gamma$-ray luminosity above 0.1\,GeV has been measured as
$L_\gamma^{\rm M15} = (5.26_{-1.16}^{+1.31})\times 10^{34}$\,erg\,s$^{-1}$. 
The spectrum is close to the power-law type with spectral index $2.84\pm 0.18$, observed up to $\sim$5~GeV \citep{zha16}. The GeV emission from a GC is interpreted as a cumulative emission from the whole population of the MSPs within the cluster. In fact, observations of nearby MSPs in the galactic plane based on the \textit{Fermi}-LAT data \citep{ab09b} allow us to estimate the conversion efficiency of the MSP's rotational energy into GeV $\gamma$-rays of $\eta_\gamma\approx 0.08$ and the average luminosity of such MSP on 
$\langle L_\gamma^{\rm MSP} \rangle = 1.44\times 10^{33}$ erg s$^{-1}$. 
Based on these average values and the MSP hypothesis of the $\gamma$-ray emission from the GCs, it is possible to estimate the number of MSPs in \m15 to be $37_{-8}^{+9}$. 
Note that the pulsations from most of the MSPs will not be observable due to their unfavourable observation angles. 
We further estimated the total rotational energy loss rate by the MSPs in \m15 on $L_{\rm rot} = L_\gamma^{\rm M15}/\eta_\gamma\approx 6.7\times 10^{35}$ erg s$^{-1}$.

It is expected that the winds of MSPs mix within the GC with the matter ejected  by the ambient RG stars. 
For the expected mass loss rates of specific RGs, estimated in the range $10^{-9}$ M$_\odot$ yr$^{-1}$ to $3\times 10^{-7}$ M$_\odot$ yr$^{-1}$ \citep{bo08,me09}, and assuming the number of RGs within \m15 to be of the order of 100 (for another GC, NGC~2808 137 RGs have been observed, \citealp{ca04}), it is possible to estimate the velocity of the mixed MSP and RG winds (see Eq.~1 in \citealp{bss16}).
Assuming that MSPs provide luminosities of the order of $L_{\rm rot}$ derived above, and that the MSP winds mix with the RG material with a mixing efficiency of 0.5, we estimate the velocity of the resulting mixed wind in the range $v_{\rm w}\approx (0.2-3.5)\times 10^8$ cm~s$^{-1}$.
The mixing efficiency is the fraction of the rotational energy loss of the pulsars that is transmitted to the stellar winds. 
It depends on the geometry of all the MSP binary systems, and is only constrained to be lower then unity.  
Note that recent estimates of the mass loss rate by RG stars in another massive GC, 47~Tuc, 
give values of about $3\times 10^{-6}$~M$_\odot$~yr$^{-1}$ \citep{mc15}, with the total content of ionized gas in 47~Tuc $\sim$0.1~M$_\odot$ (Freire et al.~2001). Quite a large amount of a neutral gas, $\approx 0.3$~M$_\odot$, has also been detected within \m15 \citep{ev03, van06}. 

MSPs can also distribute the magnetic field with their winds to the volume of the GC.  
The strength of this magnetic field can be approximated by its value at the collision region of the MSP winds (see estimates given by Eqs.~2 and~3 in \citealp{bs07}). 
For an average MSP, we assume a typical rotation period of $4$~ms and the surface magnetic field of  $3\times 10^8$\,G. 
We further assume that all MSPs in \m15 are confined within its core radius of $R_{\rm c}^{M15} = 0.43$\,pc. Then, the characteristic shock radius around the MSP in \m15 is of the order of $R_{\rm sh}\approx 2\times 10^{17}$\,cm \citep{bs07}.
The magnetization parameter, $\sigma$ which is the ratio of the Poynting flux to the relativistic particle flux, is estimated at the nebula for the winds of classical pulsars of Vela type on $\sigma\sim 0.1$ (Sefako \& de Jager~2003) and the Crab pulsar on $\sigma\sim 0.002$ \citep{kc84}.
On the other hand, if the wind terminated closer to the pulsar as expected in the case of the MSPs within GCs, the magnetization is expected to be higher, $\sigma\sim 1$  \citep{ck02}.
Using the above values of magnetization we estimate the magnetic field strength at the shock to be of the order of $B_{\rm sh}\approx (1-30)\times 10^{-6}$~G. 
In our modelling of the $\gamma$-ray emission from \m15 (see Section~\ref{sec:model}) we apply the values of the mixed winds' velocity and the magnetic field strength of the order of the ones estimated above.

\section{MAGIC observations of \m15}\label{sec:magicm15}

\begin{figure}
  \includegraphics[width=0.49\textwidth]{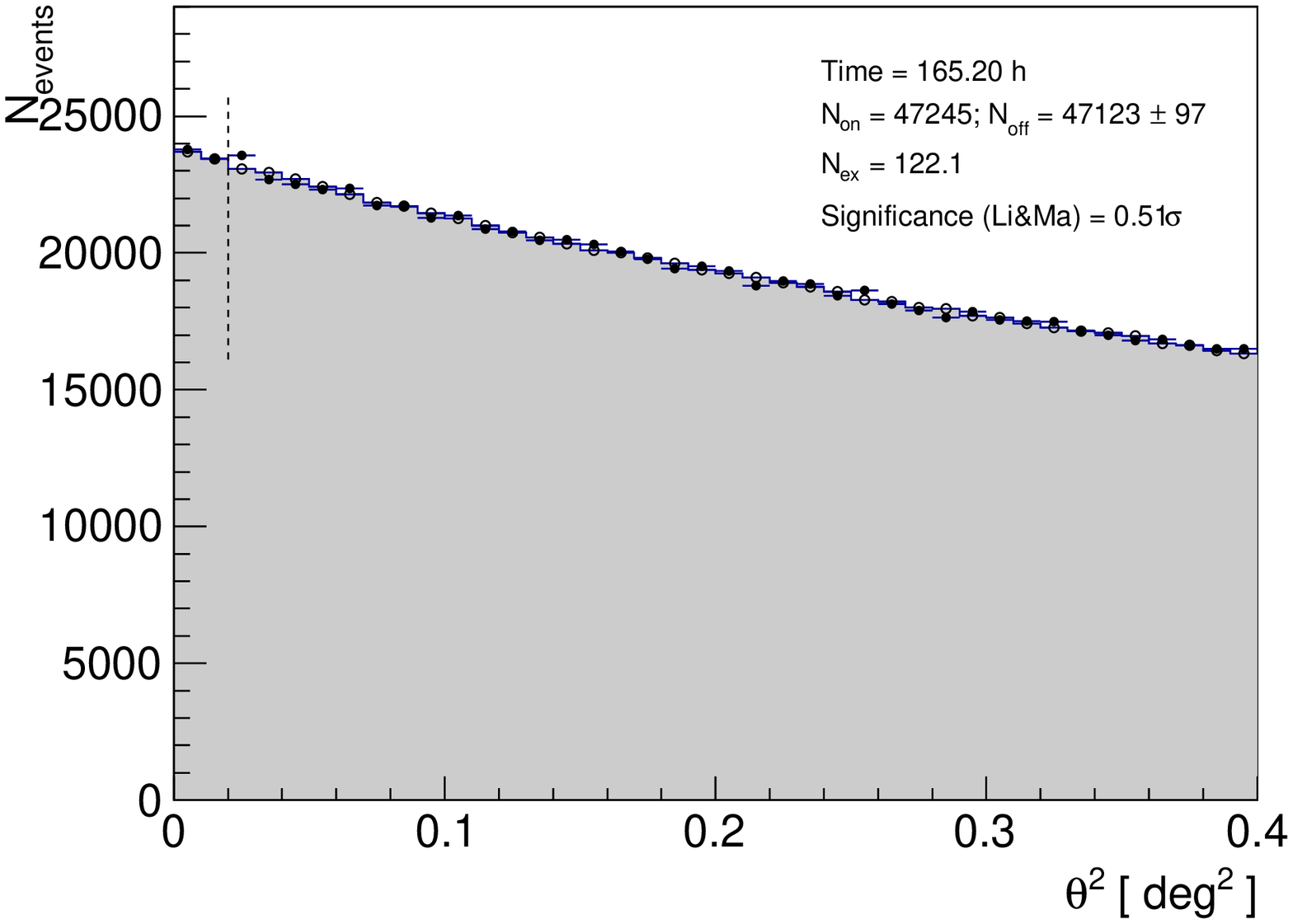}
  \includegraphics[width=0.49\textwidth]{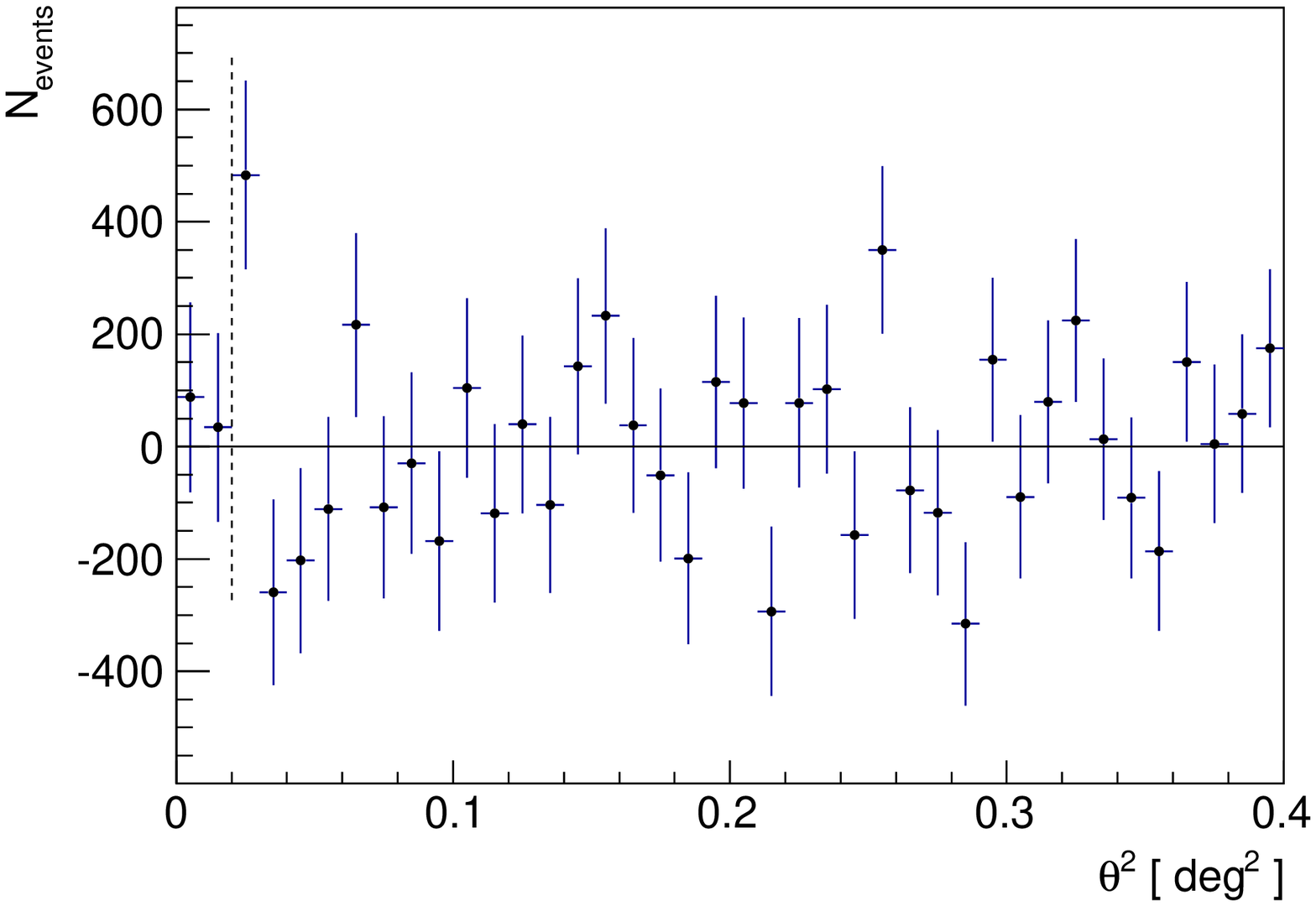}
  \caption{Top panel: distribution of the squared angular distance between the reconstructed event direction and the nominal source position (filled points) and the background estimation (shaded area and empty points).
    Bottom panel: On-Off excess distribution. 
The corresponding energy threshold (defined as the peak of the differential energy distribution for Monte Carlo 
$\gamma$ rays with a spectral index of $-2.6$) is $\sim\,75\,$GeV.
The significance of the excess was computed using equation 17 in \citet{LiMa}.}
\label{fig:th2}
\end{figure}

MAGIC is a system of two 17\,m diameter Imaging Atmospheric Cherenkov Telescopes.
The telescopes are located at the Observatorio del Roque de los Muchachos, on the Canary Island of La Palma, Spain \citep{al16a}. 
MAGIC is used for observations of $\gamma$ rays with energies between $\sim\,$50\,GeV and few tens of TeV.
The telescopes reach a sensitivity of $(0.66\pm0.03)$\% of C.U. (Crab Nebula flux) in 50 h of observations for energies above 220\,GeV \citep{al16b}. 
The angular resolution (defined as a standard deviation of a two-dimensional Gaussian distribution) at those energies is $\lesssim0.07^\circ$.

Between June 2015 and September 2016 the MAGIC telescopes performed observations of the \m15 region. 
In total 173\,hrs of data were collected, out of which 165\,hrs were selected for further analysis.
To assure a low energy threshold the data have been taken mostly at low ($\lesssim30^\circ$) zenith angles. 
The data were analyzed using the standard MAGIC analysis chain \citep{za13,al16b}. 
About one third of the selected data set was taken under non-perfect atmospheric conditions (mostly due to calima, a dust wind from the Sahara desert, which  affects part of the MAGIC data taken during summer). 
That part of the data set has been corrected using simultaneous LIDAR measurements \citep{fg15}.
 
\begin{figure}
  \includegraphics[width=0.49\textwidth]{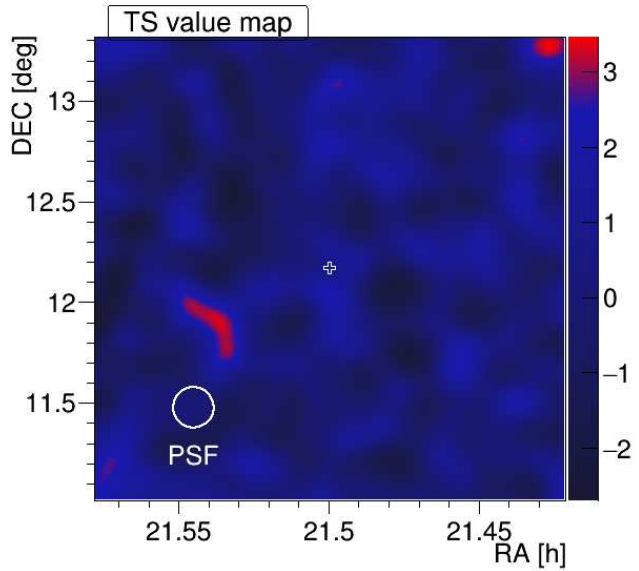}
\caption{Skymap of the \m15 observations. 
The TS value corresponds to a significance of an excess at a given location in the sky. 
The marker shows the nominal position of \m15. 
The corresponding energy threshold, $\sim 75$\,GeV, is described in Fig.~\ref{fig:th2} caption.}
\label{fig:skymap}
\end{figure}

No significant signal has been observed from the direction of \m15 (see Fig.~\ref{fig:th2}).
Also, no other significant emission is detected in the field of view covered by those observations (see Fig.~\ref{fig:skymap}). 
We compute upper limits on the flux from \m15 following the approach of \cite{ro05} using a 95\% confidence level and assuming a 30\% total systematic uncertainty on the collection area and a spectral index of $-2.6$.  In addition, we require that the upper limit on the number of excess events in each energy bin is at least 3\% of the residual background. 
This ensures that, despite the long observation time, the background-induced systematic uncertainties do not exceed the assumed systematic uncertainty (see \citealp{al16b}).
The obtained upper limit on the integral flux above 300\,GeV is equal to $3.2\times 10^{-13}\mathrm{cm^{-2}s^{-1}}$, which corresponds to $<0.26\%$ of the Crab Nebula flux. 
It is a factor of a few below the previous upper limit reported from \m15 by the H.E.S.S. Collaboration (equal to $7.2\times 10^{-13}$ cm$^{-2}$~s$^{-1}$ above 440 GeV, corresponding to 0.9\% of the Crab Nebula flux for a point like source; \citep{abr13}.
The currently most stringent constraint on the absolute TeV luminosity of a GC comes from H.E.S.S. observations of 47~Tuc.
The upper limit on the integral flux of 47~Tuc above 800 GeV is $\sim$2$\%$ of the Crab Nebula flux, which translates to a luminosity limit $6.8\times 10^{33}$~erg~s$^{-1}$ for a distance of 4 kpc \citep{ah09}.
In order to compare with the above limit on the energy flux, we compute differential flux limits and integrate them above a given energy.
MAGIC observations show that the luminosity of \m15 in the energy range 800\,GeV -- 19\,TeV is below $1.1 \times 10^{34}$~erg~s$^{-1}$, a comparable (within a factor of 2) limit to the value for 47~Tuc, despite the $\sim2.5$ times larger distance to \m15.  Moreover, the observations with
MAGIC allow us to probe the possible $\gamma$-ray emission down to lower energies without much loss of sensitivity, 
e.g. the luminosity of \m15 in the energy range 300\,GeV -- 19\,TeV derived from the MAGIC observations is below $1.6 \times 10^{34}$~erg~s$^{-1}$. The upper limits on the differential flux from \m15 are compared with the model predictions in 
Section~\ref{sec:model}. 

Relativistic leptons, ejected from the inner magnetospheres of the MSPs with a quasi mono-energetic spectrum, can comptonize the optical radiation from the GC at energies affected by the Klein-Nishina regime.

\begin{figure}
  \includegraphics[width=0.49\textwidth]{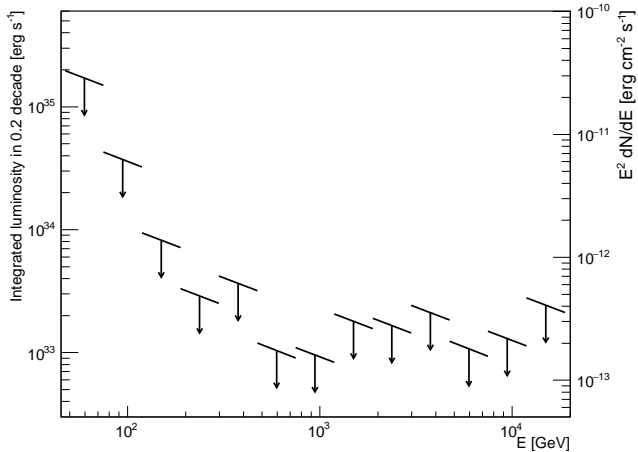}
  \caption{Upper limits on the $\gamma$-ray luminosity from the globular cluster \m15 (integrated in 0.2 decades in energy).
  The right axis shows the corresponding upper limits on the spectral energy distribution. }
\label{fig:limlum}
\end{figure}

As a result, strongly peaked $\gamma$-ray spectra are also  expected. Therefore,  we also show the upper limits on the $\gamma$-ray luminosity in narrow energy ranges 
(see Fig.~~\ref{fig:limlum}).
No hint of the emission has been seen from any of the energy bins (namely the significance was below $1.3\sigma$ in every of them).
The upper limits, integrated in 0.2 decades in energy, reach down to a level of $\sim$10$^{33}$~erg~s$^{-1}$.

\section{Theoretical expectations versus TeV $\gamma$-ray observations}\label{sec:model}

The strong upper limits on the TeV $\gamma$-ray flux from \m15 are confronted with the predictions of 
the model for the TeV $\gamma$-ray production in GCs of \cite{bss16}. 
In this model, the TeV $\gamma$ rays are produced by leptons either injected with the mono-energetic spectra directly from the inner pulsar magnetospheres or accelerated with the power-law spectra in the MSP wind collision regions. The leptons comptonize soft radiation from the huge population of low-mass stars within the GC, from the CMB, and/or optical and infrared photons from the Galactic disk. 
The model also includes the synchrotron energy losses of the leptons during their propagation in the GC. The level of produced TeV $\gamma$-ray emission in a specific GC depends on the injection rate of the leptons from the MSPs and their transport through the volume of the GC. The model considers two transport processes of the leptons, i.e. diffusion process with the Bohm diffusion prescription and the advection with the wind formed within the GC. Such a wind is expected to be composed of a mixture of the energetic MSP cumulative winds and the cumulative winds from the RGs present within the GC. The range of velocities of the GC winds can be estimated to $\sim$10$^7-10^8$~cm~s$^{-1}$, based on the known expected rate of the mass loss rate by the RGs and the energy loss rate of the MSPs within the GC (see Eq.~1 in \citealp{bss16}). 
The energy loss rate of the MSPs within \m15 is calculated for the known GeV $\gamma$-ray luminosity of the MSPs within \m15 and for the known conversion efficiency of the rotational energy of isolated MSPs to the $\gamma$-ray energy range \citep{ab09a}.
In the calculations we do not use the estimated number of MSP in \m15, but only their total luminosity measured directly by \textit{Fermi}-LAT. 
For the applied transport model of the leptons, the observations of the TeV $\gamma$-ray emission from specific GCs allow us to estimate the basic parameters characterising processes occurring in the MSP inner magnetospheres, such as the injection rate of the leptons and their energies and efficiency of the lepton acceleration in the wind regions around MSPs. 
\cite{bss16} predicted the fluxes of TeV $\gamma$ rays from \m15 for the range of parameters describing the transport of the leptons through the GC and for the two above mentioned spectra of injected leptons, i.e. mono-energetic electrons and electrons injected with a power-law spectrum. 
We use this model to interpret the results of the MAGIC observations reported in this paper. 
\footnote{
Another numerical code, which calculates the multiwavelength emission from GCs in terms of a similar model, was presented in \cite{ve05,ve08,kop13,ndi18}. This code solves the transport equation for the leptons in the GC, investigates the effects related to different diffusion models, and computes the synchrotron and IC emission from the leptons. The code has been applied for the modelling of the southern GCs such as Ter~5 and 47~Tuc, but the model predictions for \m15 are not available at present.}

\subsection{Quasi-monoenergetic leptons from MSPs}

\begin{figure*}
\vskip 5truecm
\includegraphics{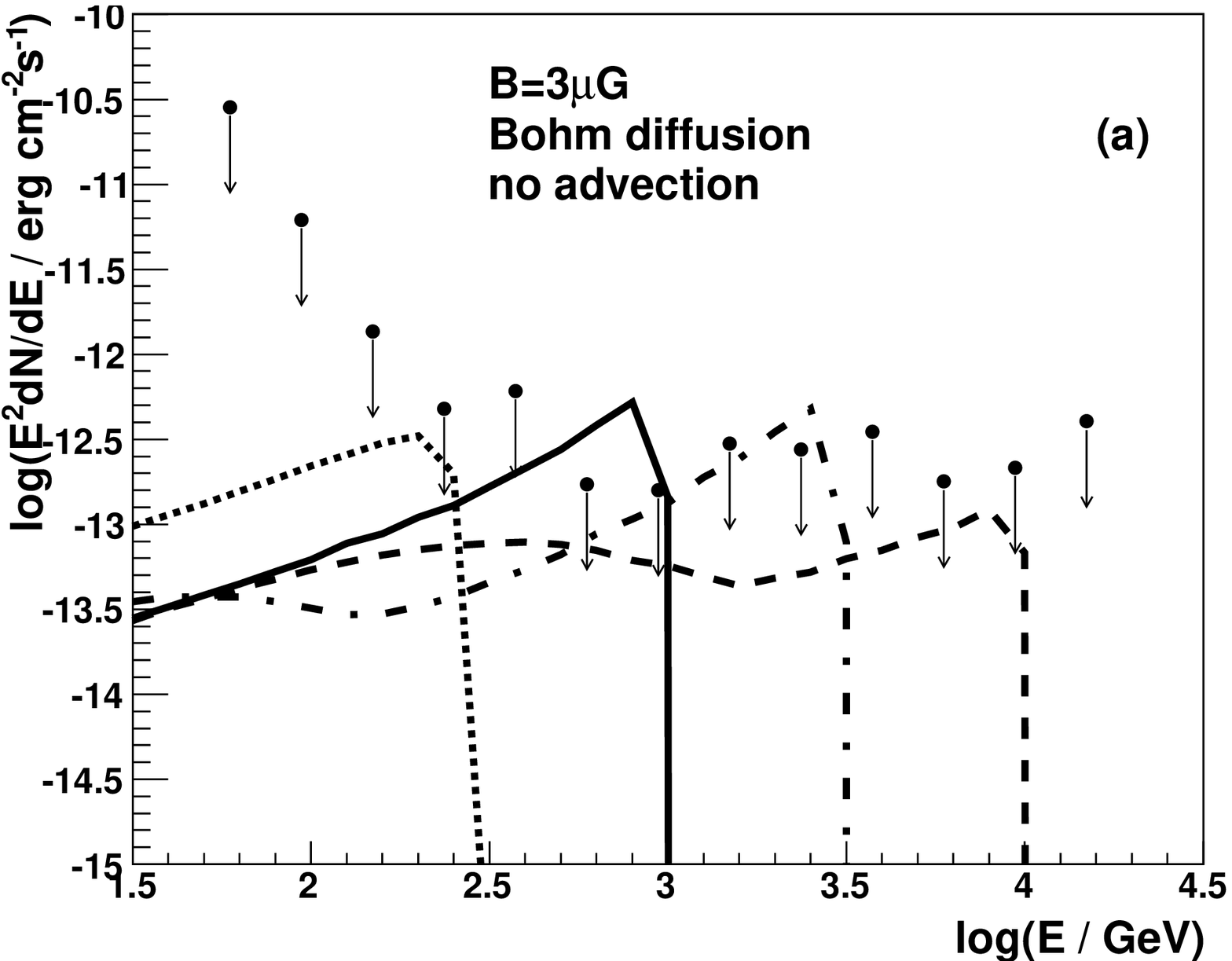}
\includegraphics{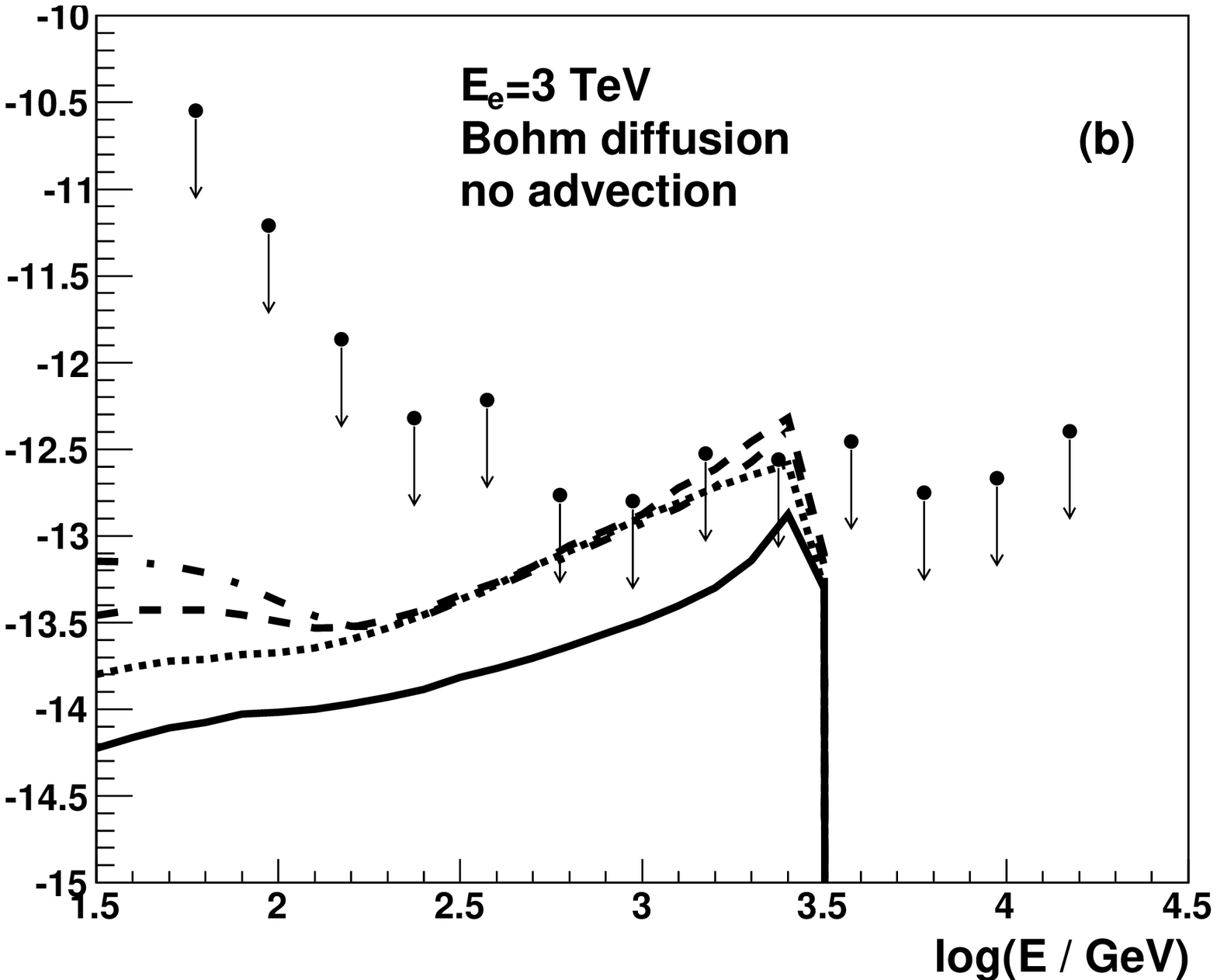}
\includegraphics{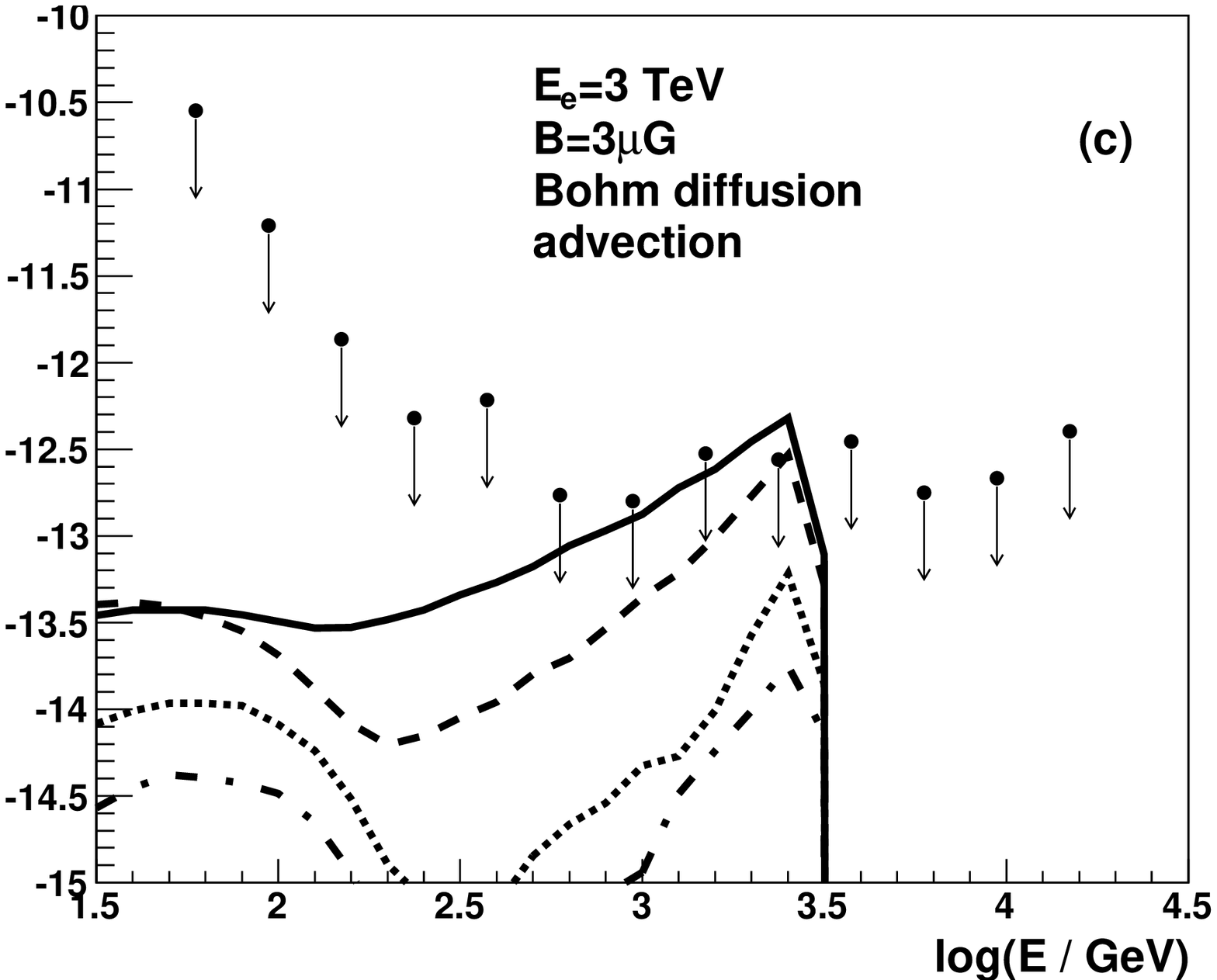}
\caption{Differential flux upper limits from 165\,h of MAGIC observations of \m15 compared with the SED produced in the case of the isotropic injection of mono-energetic leptons from the pulsars within GC \m15. We show the dependence of the IC spectra as a function of 
(a) the energy of injected leptons, $E_{\rm e} = 300$ GeV (dotted), 1 TeV (solid), 3 TeV (dot-dashed), 10 TeV (dashed), for the magnetic field strength is $B = 3\times 10^{-6}$ G and no advection; (b) on the magnetic field strength within the cluster, $B = 1\,\mu$G (dot-dashed), $3\,\mu$G (dashed), $10\,\mu$G (dotted), and $30\,\mu$G (solid), and the energy of leptons 3 TeV without advection;  (c) the advection velocity from the GC, $v_{\rm adv} = 10^7$ cm s$^{-1}$ (dashed curve),  $10^8$ cm s$^{-1}$ (dotted), $3\times 10^8$ cm s$^{-1}$ (dot-dashed), and the energy of the leptons 3\,TeV, $B = 3\times 10^{-6}$\,G and without advection (solid). It is assumed that the power injected in leptons is equal to 
1$\%$ of the rotational energy loss rate of MSPs within the GC \m15 ($L_{\rm e} = 0.01L_{\rm rot}$). 
}
\label{fig3}
\end{figure*}

We compare the differential flux upper limits, derived from MAGIC observations of \m15, with the expected TeV $\gamma$-ray spectra for a variety of parameters of the mono-energetic injection of relativistic leptons from the MSP magnetospheres (see Fig.~\ref{fig3}).
The monoenergetic spectrum considered in the model, results from assuming that all pulsars have the same potential drop (and therefore the electrons are accelerated to the same energy). 
The expected fluxes are calculated assuming that the power in relativistic leptons is equal to $\eta_{\rm e}=1\%$ of the rotational energy loss rate of the whole population of the MSPs within \m15 (i.e. $L_{\rm e} = 0.01L_{\rm rot} = 0.01L_\gamma^{\rm M15}/\eta_\gamma$), where the $\gamma$-ray power, $L_\gamma^{\rm GC} = 5.26\times 10^{34}$~erg~s$^{-1}$, has been derived based on the {\it Fermi}-LAT observations, see \citep{zha16}. The power in relativistic electrons and the observed GeV $\gamma$-ray emission can be expressed as $L_{\rm e} = \eta_{\rm e} L_{\rm rot}N_{\rm MSP}$ and $L_\gamma^{\rm GC} = \eta_\gamma L_{\rm rot}N_{\rm MSP}$ respectively. $\eta_{\rm e}$ and $\eta_\gamma$ are factors describing fractions of the MSPs rotational energy, $L_{\rm rot}$, transferred to relativistic electrons and emitted in GeV $\gamma$-rays, and $N_{\rm MSP}$ is the number of MSPs in GC. 
Note that $L_{\rm rot}\approx \langle L_\gamma^{\rm MSP}\rangle/\eta_\gamma$, where $\eta_\gamma\approx 0.08$ and 
$\langle L_\gamma^{\rm MSP} \rangle = 1.44\times 10^{33}$ erg s$^{-1}$ (the average luminosity of the MSPs) have been estimated based on the observations of nearby MSPs in the galactic field based on the \textit{Fermi}-LAT data \citep{ab09a}.
Therefore, $\eta_{\rm e} = \eta_\gamma L_{\rm e}/L_\gamma^{\rm GC}\approx 0.08L_{\rm e}/L_\gamma^{\rm GC}$. 
Then, constraints on $L_{\rm e}$ from the MAGIC observations set direct constraints on the parameter $\eta_{\rm e}$. The value of this parameter is determined by the radiation processes in the inner MSP magnetospheres and its surroundings. Its value should be predicted by different models for the high energy processes around pulsars. Note that we do not take into account the uncertainty of $L_\gamma$ when computing the limits on $\eta_{\rm e}$, since their effect is negligible compared to the spread of the limits due to the tested ranges of other model parameters. 

\begin{figure*}
\vskip 5truecm
\includegraphics{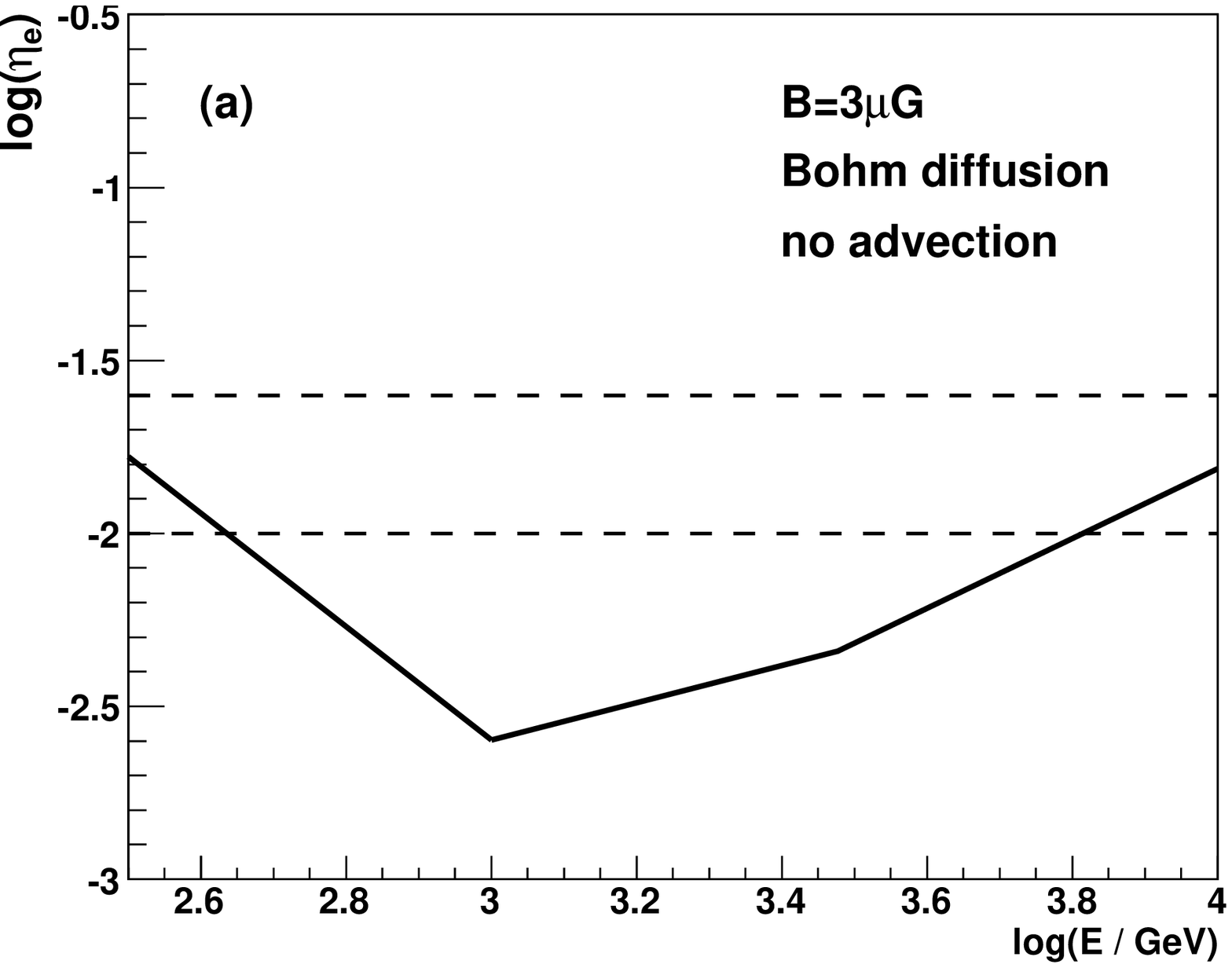}
\includegraphics{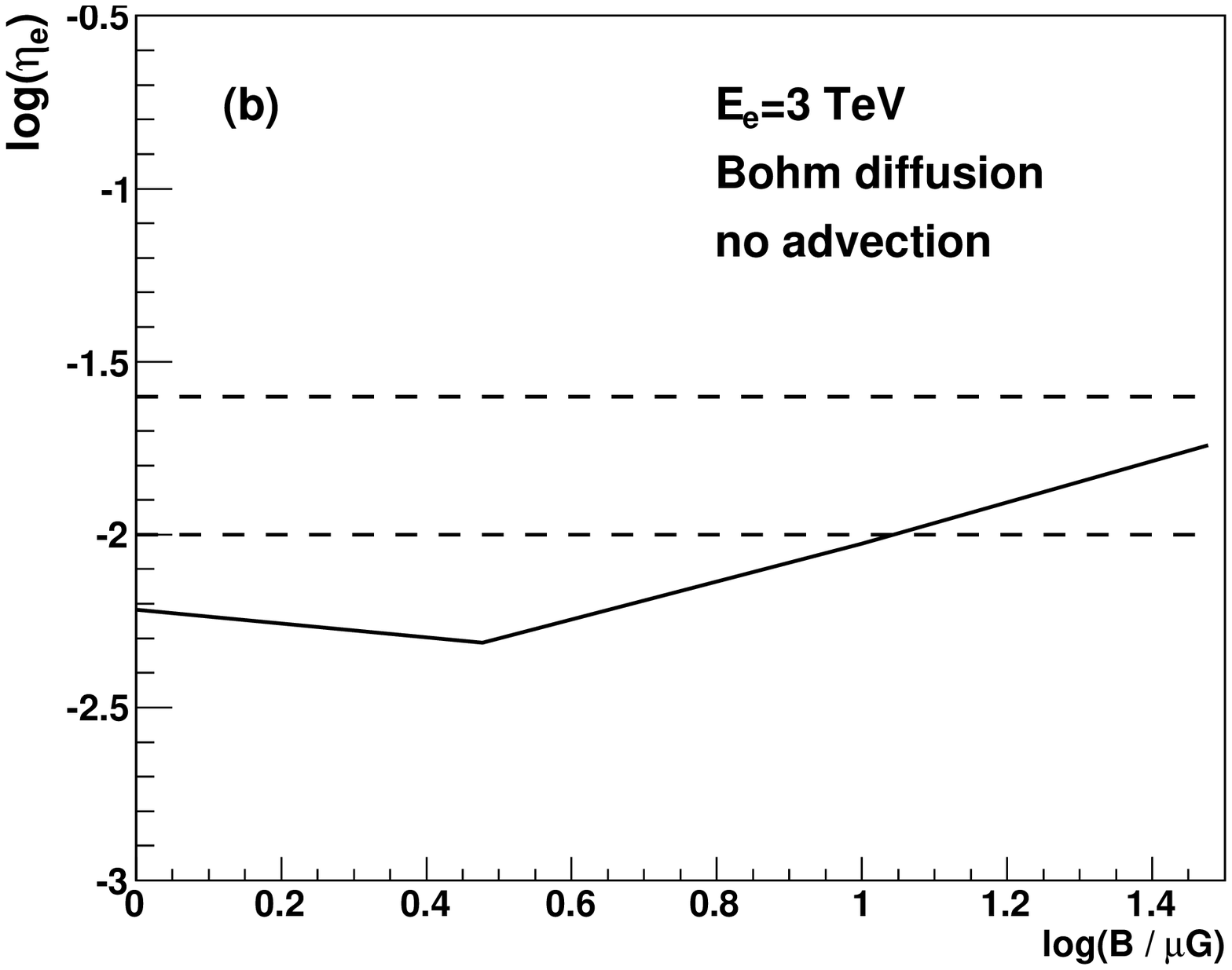}
\includegraphics{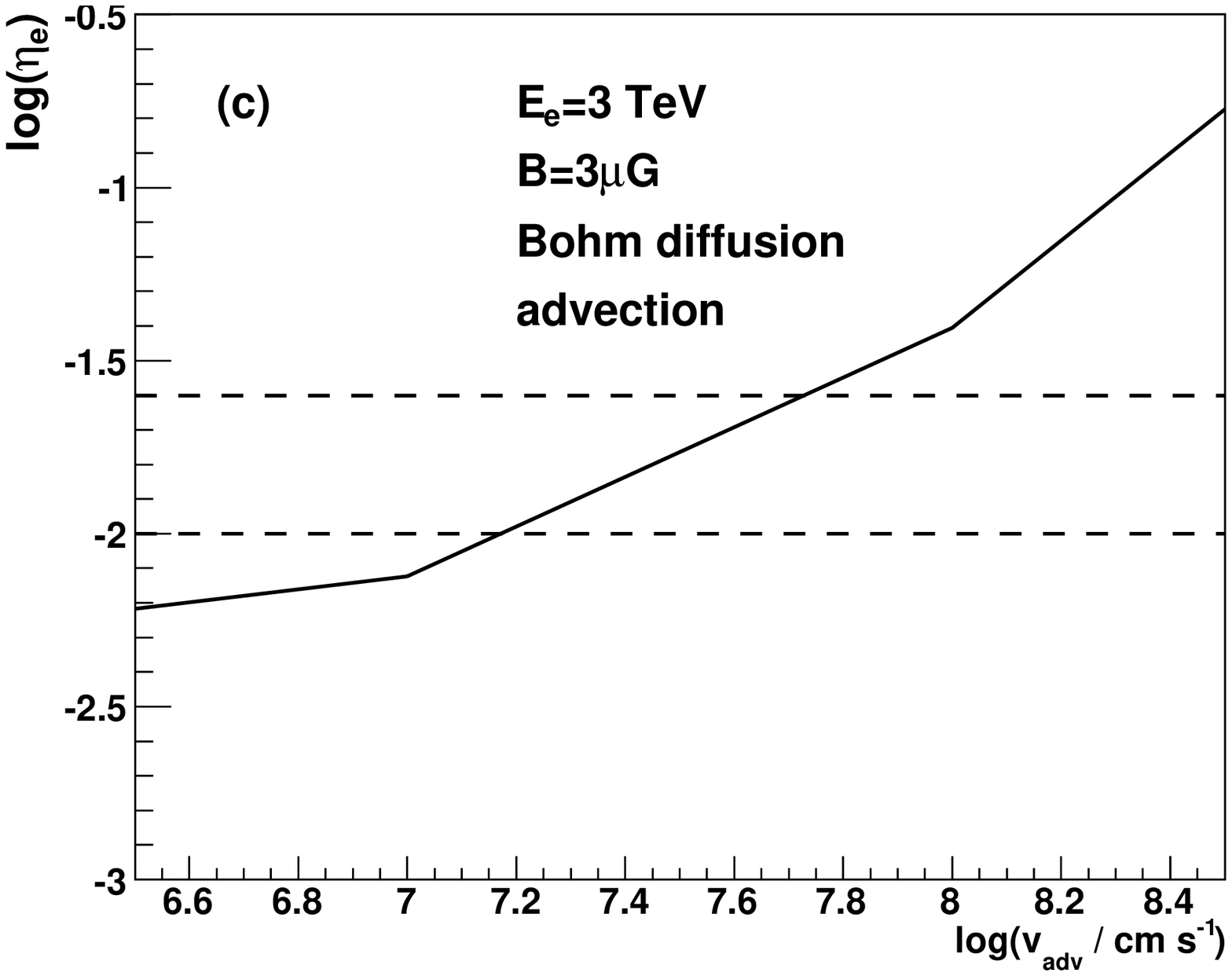}
\caption{Constraints on the coefficient, $\eta_{\rm e}$, of the conversion of the energy loss rate of the pulsars to relativistic leptons in GC \m15 in the case of injection of monoenergetic leptons from the inner pulsar magnetospheres. 
The dashed lines mark the level of the conversion efficiency expected from the pulsar models ($1-2.5\%$, see text). 
The upper limits on $\eta_{\rm e}$, obtained as a function of energy of the injected leptons, are shown in panel (a) as the solid curve. It is assumed that leptons diffuse from the GC as expected in the Bohm diffusion model and the magnetic field strenght in GC is fixed on 3\,$\mu$G. The advection of leptons is not considered. The upper limits on $\eta_{\rm e}$ obtained for different strengths of the magnetic field within GC are shown in (b) for the energy of the monoenergetic leptons equal to 3 TeV, assuming the Bohm diffusion prescription and neglecting the advection from the GC. In (c) we show the upper limits on $\eta_{\rm e}$ as a function of the advection velocity of the mixed MSP/stellar winds assuming the Bohm diffusion prescription, energy of leptons 3 TeV. }
\label{fig4}
\end{figure*}

In Fig.~\ref{fig3}a we investigate the results of such a comparison for mono-energetic leptons with energies in the range between 0.3-10~TeV, i.e. the range expected for the leptons escaping from the inner pulsar magnetospheres.
For the four considered values of lepton energies, the derived upper limits on the TeV $\gamma$-ray flux from \m15 are close or below the model predictions for the case of $L_{\rm e} = 0.01L_{\rm rot}$.
In order to be consistent with the observations, the efficiency of energy conversion from the MSPs to relativistic electrons should be below $\eta_{\rm e}\approx (0.2-2)\%$, depending on the lepton energy (see Fig.~\ref{fig4}a).
We also tested the dependence of the TeV $\gamma$-ray emission on the strength of the magnetic field within the GC, which determines the diffusion process and the synchrotron energy losses of leptons, by investigating the likely range of values 
of the random magnetic field 1-30~$\mu$G (Fig.~\ref{fig3}b). The strongest considered values of the magnetic fields have a significant effect on the production of the TeV $\gamma$-ray spectra. For strong magnetic fields, the decrease of the $\gamma$-ray flux is due to the effective synchrotron energy losses of leptons. Based on these comparisons, we set the upper limits on $\eta_{\rm e}$ on the level of $(0.5-2)\%$ for the magnetic field strengths in the range of 30\,$\mu$G -- 3\,$\mu$G, 
respectively (see Fig.~\ref{fig4}b). Finally, we compare the upper limits for the flux from \m15 with the 
TeV $\gamma$-ray spectra expected for the advection velocities in the range
$(1-30)\times 10^7$~cm~s$^{-1}$ (Fig.~\ref{fig3}c). The dependence on the wind velocity is the most critical parameter in constraining the efficiency of lepton acceleration. 
We show that the TeV $\gamma$-ray flux drops for faster winds due to inefficient scattering of soft radiation within the GC. Note also the two bump structure of the $\gamma$-ray spectra for fast winds due to the scattering of the optical radiation from the GC in the Klein-Nishina regime and the CMB in the Thomson regime. 
If the velocity of the advection wind could be constrained, more stringent limits on the energy conversion from MSPs to relativistic electrons would be possible. 
In particular, large interstellar matter density inside the cluster would slow down the winds. 
We compute the limiting density of the matter as:
\begin{equation}
\rho_{\rm RG} = \dot{M}_{\rm RG} / (4\pi R_{\rm RG}^2 v_{\rm adv}) \approx 0.03 \dot{M}_{-5} / (4\pi R_{\rm pc}^2 v_8) \mathrm{[cm^{-3}]},
\end{equation}
where $\dot{M}_{\rm RG}=\dot{M}_{-5} 10^{-5} M_\odot/{\rm yr}$ is the total mass loss rate of all the RGs in within the radius $R_{\rm RG} = R_{\rm pc}$\,pc of the \m15 and $v_{\rm adv}=v_{8} 10^8 \mathrm{cm\,s^{-1}}$ is the advection speed. 
If the density of the matter in the cluster exceeds $\rho_{\rm RG}$, the advection with speed of $v_{\rm adv}$ cannot be sustained, and the obtained limits on $\eta_{\rm e}$ become more restrictive. 
Therefore, in the case of no advection, the upper limit on $\eta_{\rm e}$ becomes very restrictive, with $\eta_{\rm e} \lesssim\,0.6\%$, for leptons with energies in the range $\sim 0.6-4$~TeV and the magnetic fields within GC below 5\,$\mu$G.

It should be noted that in the above calculations we have simplistically considered that all the MSP in M15 have the same parameters.
  Differences from one MSP to another are to be expected, however are very difficult to quantify.
  If the emission would be divided between $N$ MSPs of different parameters and peaking in different energy ranges the sensitivity for detection of such emission will be worse by a factor up to $\sqrt{N}$.
  However, as the number of peaks increases, they will start to heavily overlap (as can be seen from Fig.~\ref{fig3}a this would already be the case of $\sim 7$ peaks). 
  Moreover, not all the MSP will contribute with the same fraction of the emission, and a single dominating source can be responsible for most of the emission.
  GeV emission of a GC dominated by a single pulsar is the case for J1823-3021A MSP \citep{fre11} in NGC 6624, and to the lesser extend also PSR B1821-24 in M28 \citep{joh13}. 
We conclude that the conservative limits should be $\sim\sqrt{7}$ times larger than in Fig.~\ref{fig4}, i.e.  $\eta_{\rm e} \lesssim\,1.6\%$ in the broad range of the assumed parameters.

It is interesting to confront the constraints on $\eta_{\rm e}$, obtained from the MAGIC observations of \m15, with the expectation of some models for the acceleration and radiation processes within the MSP magnetosphere and their vicinity. 
For example, based on the fully 3D general-relativistic polar cap pulsar model, \cite{ve05} estimated the values of the parameter $\eta_{\rm e}$ to be $1\% - 2.5\%$ and $\eta_\gamma$ to be $2\% - 9\%$ for the case of  PSR J0437-4715. These estimates have been generally confirmed in the analysis of the $\gamma$-ray emission from the population of MSPs in 47~Tuc, 
$\eta_{\rm e}\sim2\%$ and $\eta_\gamma\sim 7\%$ \citep{ve08}. The estimates of $\eta_\gamma\sim 10\%$ are consistent with other modelling of processes in the pulsar's inner magnetospheres based on a space charge-limited outflow polar cap model (e.g. \citealp{ha02}), the outer gap model \citep{ta10}, and with the estimates based on the {\it Fermi}-LAT observations of the population of MSPs (\citealp{ab09a,ab09b}, see also Fig.~9 in the pulsar catalogue by \citealp{ab13}). The upper limits on 
$\eta_{\rm e}$ derived in Fig.~\ref{fig4} are, for most of the parameter space, below the values expected from modelling of the processes in the pulsar magnetospheres mentioned above. However, the predictions of the pulsar models might still be consistent with the MAGIC observations provided that the mixed pulsar/stellar wind velocity is $\gtrsim\,1.5\times 10^7$~cm~s$^{-1}$ for $\eta_{\rm e} = 1\%$ and  $\gtrsim\,6\times 10^7$~cm~s$^{-1}$ for $\eta_{\rm e} = 2.5\%$  (see Fig.~\ref{fig4}c). Note however that a quite large amount of a neutral gas has been discovered within \m15 (see Sect.~2). This is difficult to explain in the case of efficient removal of the matter from the volume of this GC as a result of a very fast wind from the GC. Therefore, we conclude that the present models for the high energy processes in the inner pulsar magnetospheres are in conflict with the constraints obtained here on the injection rate of the leptons from the inner magnetospheres of the MSPs within \m15 provided that the RG winds are not able to remove very efficiently the TeV leptons from the volume of the GC.
The presence of such fast winds is not supported by the observations of substantial diffusive matter within \m15.

The parameter $\eta_{\rm e}$ can be related to the parameter $\sigma$ which characterises the form of energy injected from the inner pulsar magnetosphere. $\sigma$, defined as the ratio of the Poynting flux to the particle flux, can be related to the energy carried by the magnetic field, $L_{\rm B}$, and relativistic particles, $L_{\rm e}$, provided that the surface area of these two flows is the same, i.e. $\sigma = F_{\rm P}/F_{\rm k} = 
L_{\rm B}S_{\rm k}/L_{\rm k}S_{\rm P} = L_{\rm B}/L_{\rm e}$ for $S_{\rm p} = S_{\rm k}$. 
$L_{\rm B}$ and $L_{\rm e}$ are related to each other as $L_{\rm B} = L_{\rm rot} - L_{\rm e}$. The possible contribution from heavy ions to the energy loss rate is neglected, although it has been proposed to be also important, e.g. \cite{ga94,co17}. $\sigma$ is related to the parameter $\eta_{\rm e}$ in the following way, $\sigma = (L_{\rm rot} - L_{\rm e})/L_{\rm e} = \eta_{\rm e}^{-1} - 1$. For classical pulsars, $\sigma$ is expected to be of the order of 
$\sim$10$^{4}$ close to the light cylinder radius (e.g. \citealp{che86,ar08}).  
The MAGIC constraints on $\eta_{\rm e}$ (see Fig.~\ref{fig4}), in the case of a relatively slow advection process of leptons from \m15, allow us already to constrain $\sigma$ to be  $\gtrsim$200. Therefore, the value of the magnetization parameter 
for the MSPs at their light cylinder radius is also expected to be as large as in the case of classical pulsars.

\subsection{Leptons with a power-law spectrum}

\begin{figure*}
\vskip 5truecm
\includegraphics{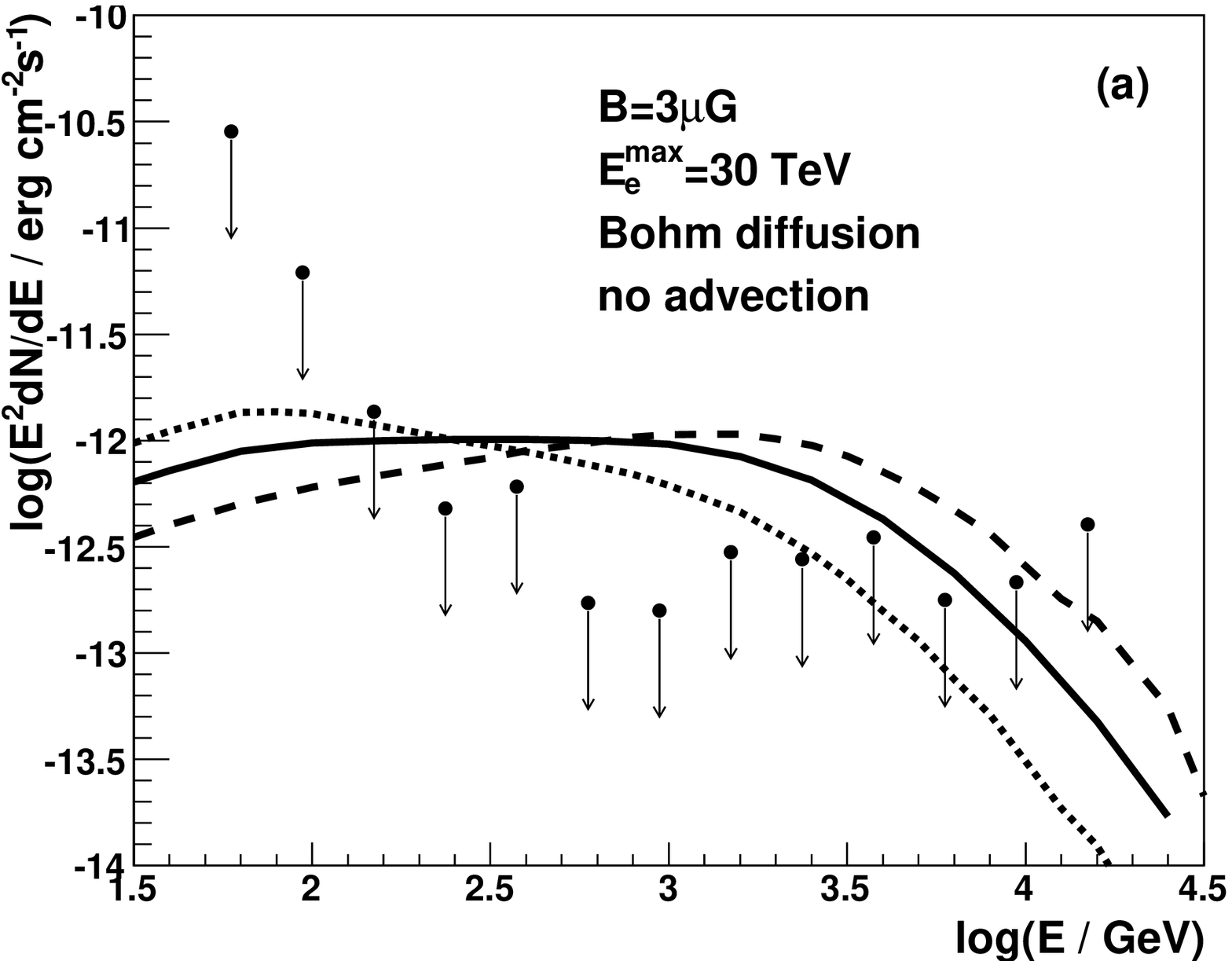}
\includegraphics{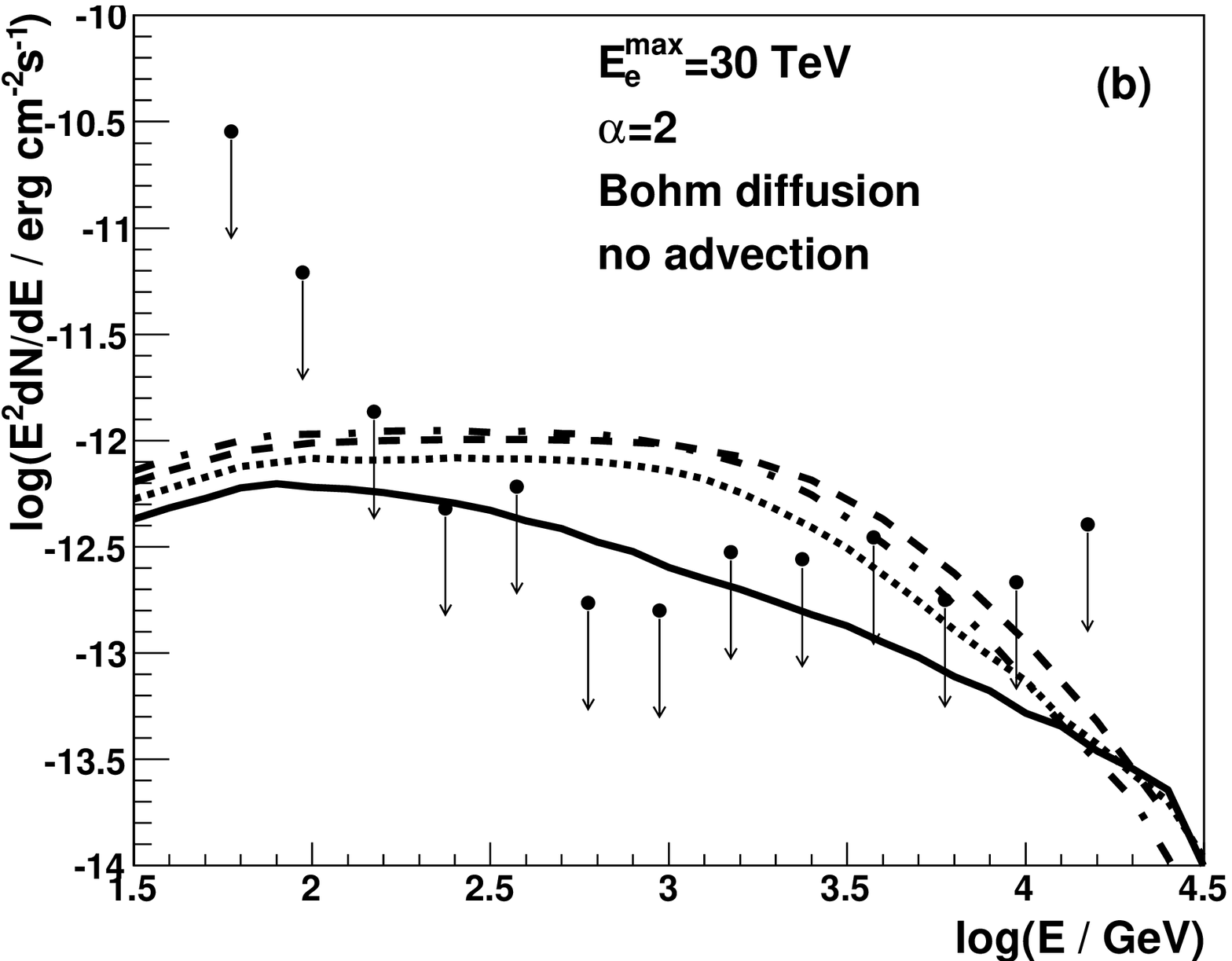}
\includegraphics{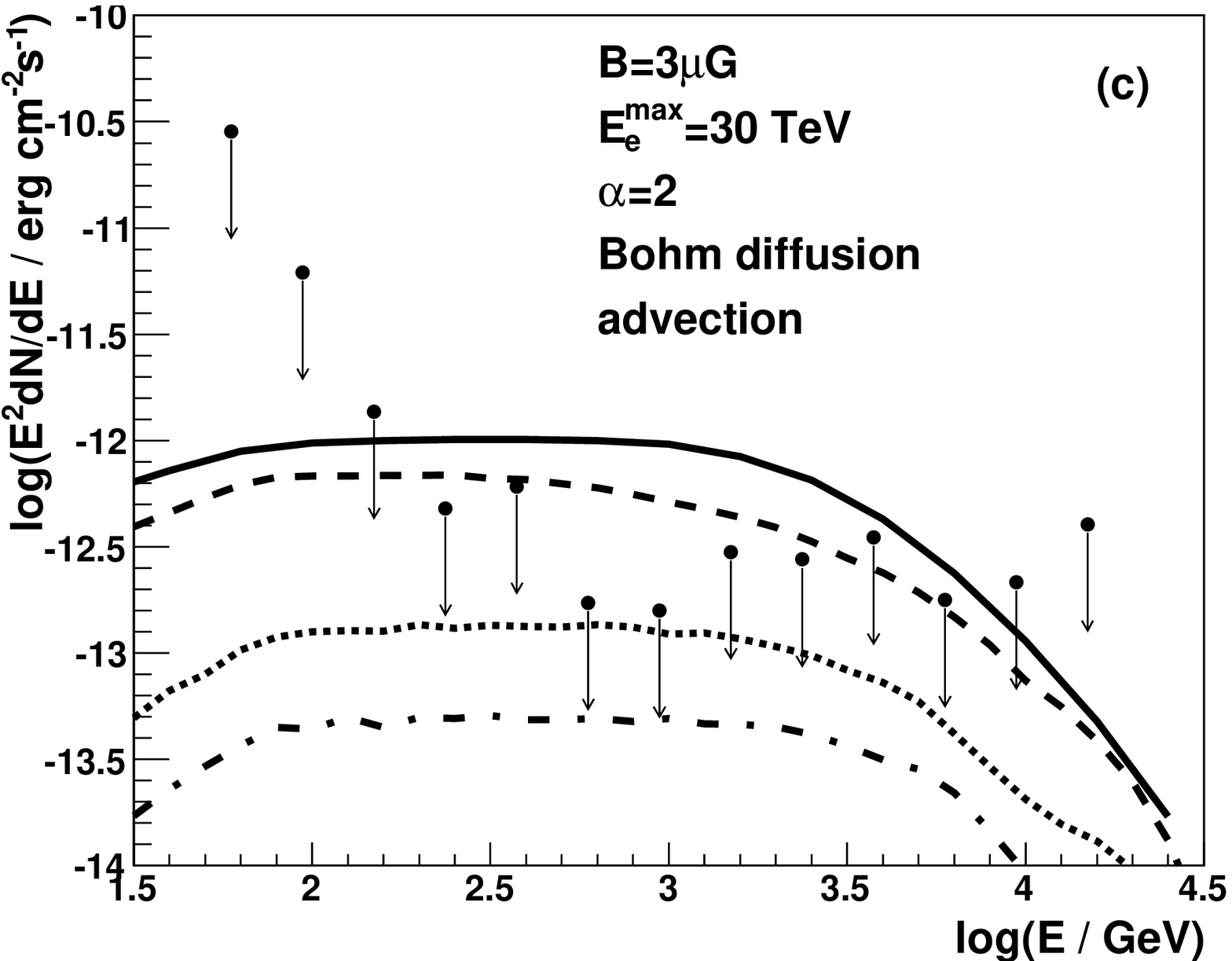}
\caption{
The differential flux upper limits from 165h of \m15 observations (points) compared with the SED produced in the case of the isotropic injection of the leptons with the power-law spectrum from the pulsars within GC \m15. We show the dependence of the IC spectra as a function of (a) the spectral index of injected leptons, $\alpha = 1.5$  (dashed), 2. (solid), 2.5 (dotted), and the cut-off energy in the spectrum at 30 TeV, the magnetic field strength $B = 3\,\mu$G in the case of no advection; (b) on the magnetic field strength within the cluster, $B = 1\,\mu$G (dot-dashed), $3\,\mu$G (dashed), $10\,\mu$G (dotted), and $30\,\mu$G (solid), and spectral index of leptons equal to $\alpha = 2$, the cut-off energy in the lepton spectrum at 30 TeV,  and without advection;  (c) the advection velocity from the GC, $v_{\rm adv} = 10^7$ cm s$^{-1}$ (dashed curve),  $10^8$ cm s$^{-1}$ (dotted), 
$3\times 10^8$ cm s$^{-1}$ (dot-dashed), and without advection (solid), the power-law spectrum of leptons with the spectral index equal to 2 and the cut-off at 30 TeV, and the magnetic field strength equal to $B = 3\,\mu$G. 
It is assumed that the power in injected leptons is equal to 10$\%$ of the rotational energy loss rate of MSPs within the GC \m15. }
\label{fig5}
\end{figure*}

\begin{figure*}
\vskip 5truecm
\includegraphics{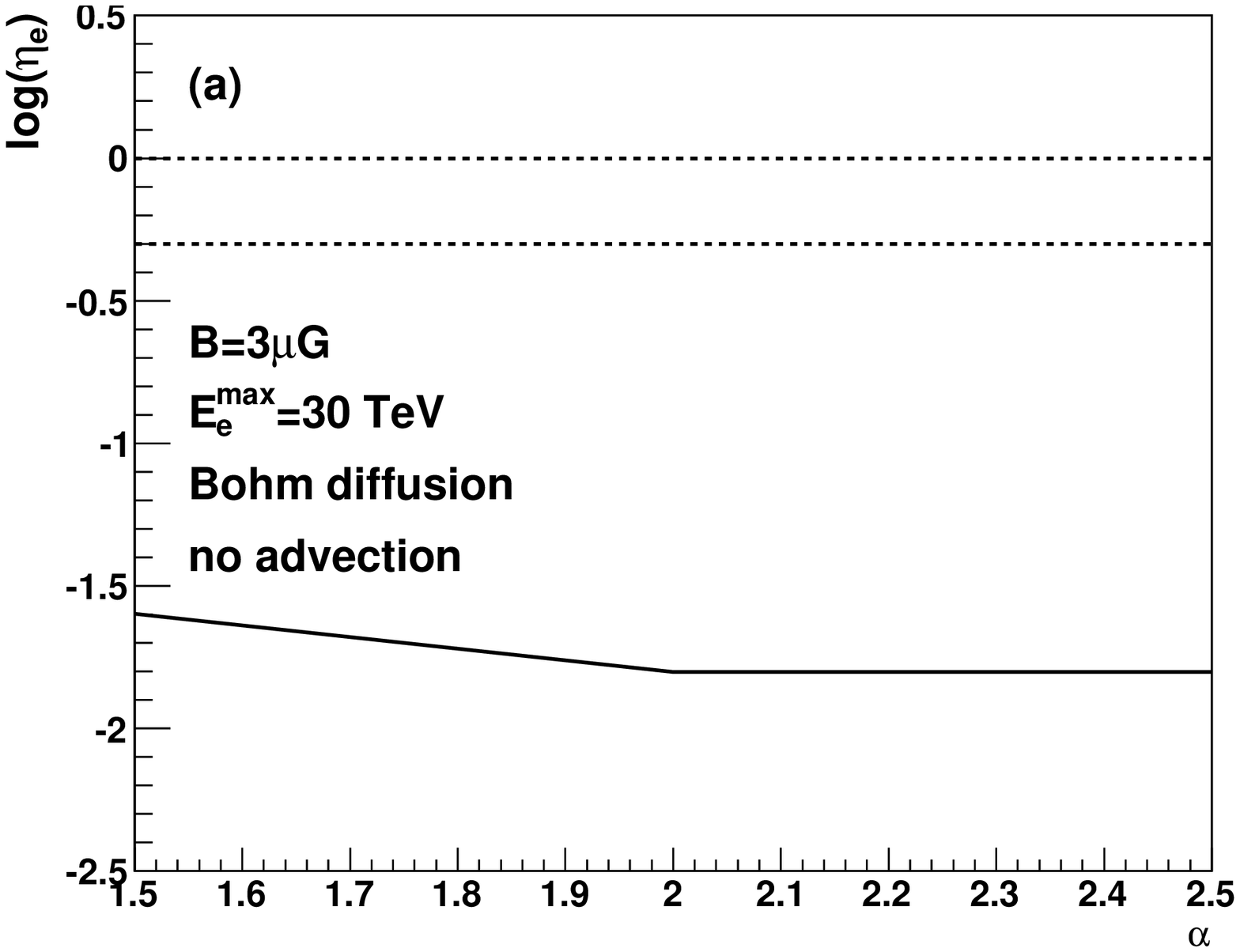}
\includegraphics{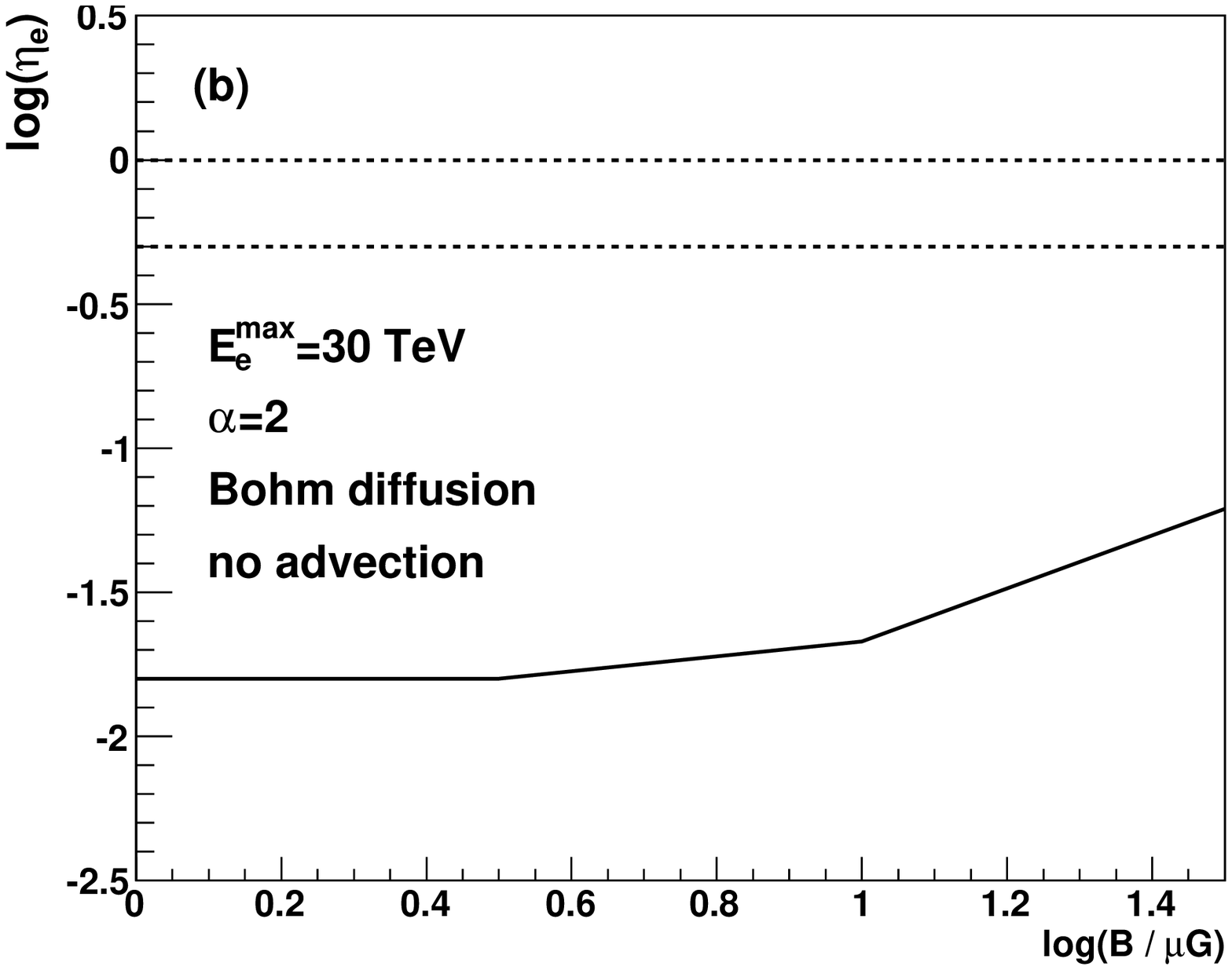}
\includegraphics{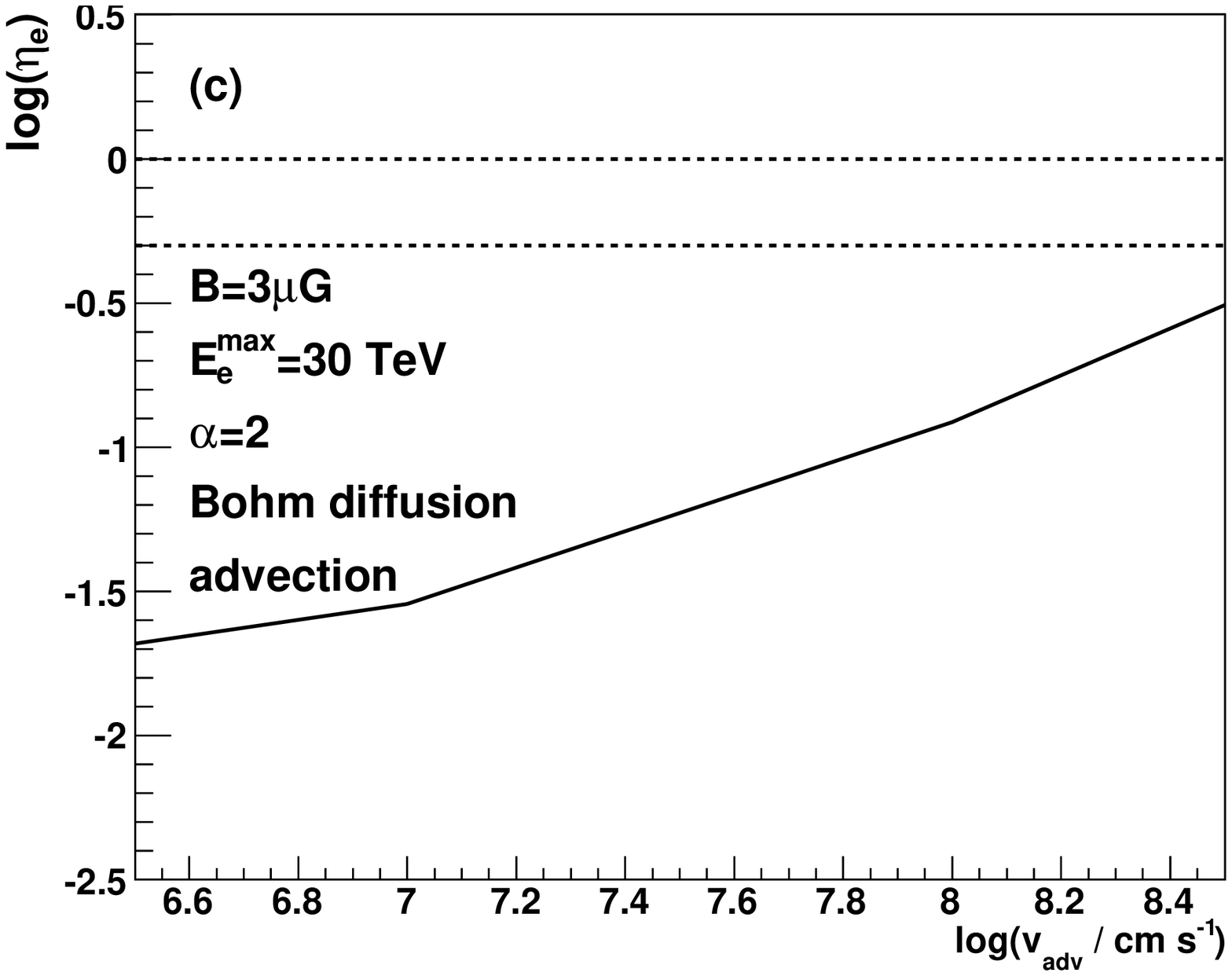}
\caption{
Constraints on the coefficient, $\eta_{\rm e}$, of the conversion of the energy loss rate of the pulsars to relativistic leptons in the GC \m15.
in the case of injection of leptons with the power-law spectrum from the pulsar wind collision regions (solid curve). 
The upper limits on $\eta_{\rm e}$ are obtained for the specific propagation and injection models corresponding to the figures (a), (b), and (c) as shown in Fig.~\ref{fig5}. The dotted lines mark the level of the energy conversion efficiency from the pulsars to leptons equal to $50\%$ and $100\%$.}
\label{fig6}
\end{figure*}

\cite{bss16} also considered the case of injection of leptons with power-law spectra from the MSP wind regions.
We have compared the results of calculations of the TeV $\gamma$-ray spectra produced in the case of the power-law injection model with the upper limits on \m15 obtained from the MAGIC observations (see Fig.~\ref{fig5}).
The dependence on the TeV $\gamma$-ray emission on the spectral index of the leptons, the magnetic field strength within \m15, and the advection velocity of the leptons from the GC are investigated. From confrontation of these calculations with the observed upper limits we derive upper limits on the acceleration efficiency of the leptons with a power-law spectrum (Fig.~\ref{fig6}).
The upper limits on $\eta_{\rm e}$ are below $\sim$2$\times 10^{-2}$ for negligible advection process (velocities below $v_{\rm adv}\sim 10^7$~cm~s$^{-1}$)  
and below $\sim$0.3 for fast advection process ($v_{\rm adv}\sim$3$\times 10^8$~cm~s$^{-1}$) and reasonable values of the magnetic field strength within the GC ($B < 10\,\mu$G). 
The values of the parameter $\sigma$, corresponding to those acceleration velocities are estimated to be $\sigma_{\rm MSP} > 49$ and $\sigma_{\rm MSP} > 2.3$ respectively, for the above mentioned conditions. 
However, for stronger fields these limits increase reaching the values of $\sim$0.8 for $B = 30\,\mu$G. 
Note that the value of $\sigma$ parameter in the nebulae around classical pulsars have been estimated to 0.002 (corresponding to $\eta_{\rm e} = 0.998$) for the Crab Nebula (e.g. \citealp{rg74,kc84}) and $\sim$0.1 ($\eta_{\rm e} = 0.909$) for the nebula around the Vela pulsar \citep{se03}. In the case of the winds, which are forced to terminate closer to the pulsar, the values of $\sigma$ have been calculated to be of the order of unity ($\eta_{\rm e} = 0.5$) \citep{ck02}. The lower limits on the $\sigma$ parameter, constrained for the winds around MSPs within \m15, are clearly above the values expected for nebulae around classical pulsars.  
Based on the above observations and theoretical calculations, we conclude that pulsar wind regions around MSPs are not able to accelerate leptons with the rates similar to those observed in the case of classical pulsars.

A few explanations of the above derived constraints on the efficiency of lepton acceleration in the wind regions of the MSPs can be elaborated. 
The low values of $\eta_{\rm e}$ might indicate that the interaction of the MSP winds with a large amount of small scale winds around classical stars (present within the GC) or with the MSP winds between themselves disrupt the wind structure preventing efficient acceleration of particles. Then, the inner regions of the pulsar winds, at distances comparable to the typical distance between the stars which is $\sim$10$^{17}$~cm in the case of MSPs in \m15, might not be able to convert efficiently energy from the magnetic field into relativistic particles.
Another possible explanation can be related to the effectiveness of removal of the wind matter from the GC. If the mixed MSP/RG winds move faster than estimated above, then the upper limits derived in our modelling from the TeV $\gamma$-ray observations of \m15 become much less restrictive. This explanation seems unlikely in the context of a relatively large amount of distributed matter observed in \m15. Finally, the low effectiveness of lepton acceleration can be related to the assumption on the wind composition around MSPs. In the case of MSPs, the winds can be energetically dominated by hadrons.
In fact, the conditions and the content of the matter at the surfaces of the classical pulsars and the MSPs within GCs differ significantly. The surface of classical pulsars is expected to contain heavy nuclei which might be strongly bounded to the surface in the strong magnetic field (Usov \& Melrose~1995). On the other hand, the mildly magnetized magnetospheres of the MSPs within GCs are expected to be composed of the matter accreted from the atmospheres of low mass main sequence stars (i.e. mainly hydrogen and helium). Therefore, the composition of particles, which dominate energetically in the neutron star's magnetosphere, can be different. In the case of classical pulsars, leptons dominate since iron nuclei are strongly bounded to the surface. Then, the pulsar radiation processes might be well described in terms of the so-called slot gap
model (Arons~1983) or pair-starved polar cap model (see Ruderman \& Sutherland~1975, Muslinov \& Harding~2003). In the case of MSPs, light hadrons dominate since they are not strongly bounded to the neutron star surface. Then, the pulsar energy loss rate is carried mainly by energetic hadrons. In this case leptons contribute only partially to the energy loss rate of the pulsar. Then, the high energy emission from the MSPs can be better explained in terms of so called space charge limited outflow model (Arons \& Scharlemann~1979). 

We suggest that the acceleration/radiation processes within the atmospheres of classical $\gamma$-ray pulsars and the MSPs are different. Note that up to now no isolated TeV $\gamma$-ray Pulsar Wind Nebula around a MSP has been firmly detected \citep{ahn17,hl18}. This, combined with the limits derived by the MAGIC observations of \m15, suggests that the winds of the MSPs might not accelerate leptons as efficiently as the winds around classical pulsars.

\section{Conclusion}\label{sec:conc}

Based on the deep observations of the globular cluster \m15 with the MAGIC telescopes, we obtained the most constraining upper limit 
on the TeV $\gamma$-ray emission of $0.26\%$ Crab flux above 300 GeV. This upper limit has been compared with the predictions of the IC emission model in which leptons are accelerated in the inner magnetospheres (with a quasi-monoenergetic spectrum) or in the wind regions (with a power-law spectrum) of the MSPs (see Bednarek et al.~2016).

Based on the comparison of the MAGIC upper limits on \m15 with the predictions of the model, we constrain the efficiency of lepton injection from the inner magnetospheres of MSPs and their vicinity. We conclude that the injection rate of leptons from the inner pulsar magnetospheres is below the rate expected from MSPs (e.g. Venter \& de Jager~2005, Venter \& de Jager~2008) for the expected range of the free parameters describing the IC model. We also constrained the injection rate of leptons with a power-law spectrum from the MSP wind regions. This injection rate of the leptons is also clearly below the values estimated for the injection from the wind regions around classical pulsars. 

We discuss possible explanation of these very low injection rates  from the MSPs within \m15. The limits become less restrictive in the case of a fast advection of the leptons from the region of the GC with the velocity of the wind formed as a result of the mixture of the MSP winds with the winds from the population of the RGs within \m15. This explanation is difficult to reconcile with the presence of a relatively large quantity of neutral gas within \m15 (e.g. Evans et al.~2003, van Loon et al.~2006).
An important difference between the wind regions around classical pulsars and those around MSPs within GCs is related to the presence of a large number of local, low mass stellar winds. Also the conditions at the termination shocks of the winds around
these two types of pulsars can significantly differ since the winds around MSPs can collide between themselves.
Therefore, leptons might not be able to reach  large energies in the case of such earlier disrupted winds around MSPs. This suggests that leptons are only effectively accelerated in the pulsar wind regions which are terminated at relatively large distances from the pulsars.

We also speculate that the wind regions around MSPs and classical pulsars might have different compositions which results in a different mechanism for acceleration and radiation of the leptons in the vicinity of the pulsars.
In fact, it is expected that the composition of matter on the surface of neutron stars, which are observed as classical pulsars and MSPs, differs significantly between MSPs and classical pulsars. MSPs are expected to contain light nuclei (hydrogen, helium) on the surface which are not strongly bound to the NS surface. On the other hand, the atmosphere of the classical pulsar is mainly composed of iron nuclei which might be strongly bound to the surface in the super-strong magnetic field. Then, the energy lost by the pulsar could be mainly carried out in the form of relativistic leptons in the case of classical pulsars (i.e. in terms of the extended polar cap model) and in the form of relativistic light nuclei in the case of MSPs (i.e. in terms of the space-charge limited outflow model). As a result, the injection rate of leptons from MSPs should be expected to be clearly lower than the injection rate of leptons from classical pulsars, in consistency with the present MAGIC observations of the GC \m15.  

\section*{Acknowledgements}
%
%
We would like to thank the Referee for useful comments and 
the Instituto de Astrof\'{\i}sica de Canarias for the excellent working conditions at the Observatorio del Roque de los Muchachos in La Palma. The financial support of the German BMBF and MPG, the Italian INFN and INAF, the Swiss National Fund SNF, the ERDF under the Spanish MINECO (FPA2015-69818-P, FPA2012-36668, FPA2015-68378-P, FPA2015-69210-C6-2-R, FPA2015-69210-C6-4-R, FPA2015-69210-C6-6-R, AYA2015-71042-P, AYA2016-76012-C3-1-P, ESP2015-71662-C2-2-P, FPA2017‐90566‐REDC), the Indian Department of Atomic Energy and the Japanese JSPS and MEXT is gratefully acknowledged. This work was also supported by the Spanish Centro de Excelencia ``Severo Ochoa'' SEV-2016-0588 and SEV-2015-0548, and Unidad de Excelencia ``Mar\'{\i}a de Maeztu'' MDM-2014-0369, by the Croatian Science Foundation (HrZZ) Project IP-2016-06-9782 and the University of Rijeka Project 13.12.1.3.02, by the DFG Collaborative Research Centers SFB823/C4 and SFB876/C3, the Polish National Research Centre grant UMO-2016/22/M/ST9/00382 and by the Brazilian MCTIC, CNPq and FAPERJ.
This work is also supported by the grant through the Polish National Research Centre No. 2014/15/B/ST9/04043.


\vspace*{0.5cm}
\noindent
Affiliations:\\
$^{1}$ {Inst. de Astrof\'isica de Canarias, E-38200 La Laguna, and Universidad de La Laguna, Dpto. Astrof\'isica, E-38206 La Laguna, Tenerife, Spain} \\
$^{2}$ {Universit\`a di Udine, and INFN Trieste, I-33100 Udine, Italy} \\
$^{3}$ {National Institute for Astrophysics (INAF), I-00136 Rome, Italy} \\
$^{4}$ {ETH Zurich, CH-8093 Zurich, Switzerland} \\
$^{5}$ {Technische Universit\"at Dortmund, D-44221 Dortmund, Germany} \\
$^{6}$ {Croatian MAGIC Consortium: University of Rijeka, 51000 Rijeka; University of Split - FESB, 21000 Split; University of Zagreb - FER, 10000 Zagreb; University of Osijek, 31000 Osijek; Rudjer Boskovic Institute, 10000 Zagreb, Croatia} \\
$^{7}$ {Saha Institute of Nuclear Physics, HBNI, 1/AF Bidhannagar, Salt Lake, Sector-1, Kolkata 700064, India} \\
$^{8}$ {Centro Brasileiro de Pesquisas F\'isicas (CBPF), 22290-180 URCA, Rio de Janeiro (RJ), Brasil} \\
$^{9}$ {Unidad de Part\'iculas y Cosmolog\'ia (UPARCOS), Universidad Complutense, E-28040 Madrid, Spain} \\
$^{10}$ {University of \L\'od\'z, Department of Astrophysics, PL-90236 \L\'od\'z, Poland} \\
$^{11}$ {Deutsches Elektronen-Synchrotron (DESY), D-15738 Zeuthen, Germany} \\
$^{12}$ {Istituto Nazionale Fisica Nucleare (INFN), 00044 Frascati (Roma) Italy} \\
$^{13}$ {Max-Planck-Institut f\"ur Physik, D-80805 M\"unchen, Germany} \\
$^{14}$ {Institut de F\'isica d'Altes Energies (IFAE), The Barcelona Institute of Science and Technology (BIST), E-08193 Bellaterra (Barcelona), Spain} \\
$^{15}$ {Universit\`a di Siena and INFN Pisa, I-53100 Siena, Italy} \\
$^{16}$ {Universit\`a di Padova and INFN, I-35131 Padova, Italy} \\
$^{17}$ {Universit\`a di Pisa, and INFN Pisa, I-56126 Pisa, Italy} \\
$^{18}$ {Universit\"at W\"urzburg, D-97074 W\"urzburg, Germany} \\
$^{19}$ {Finnish MAGIC Consortium: Tuorla Observatory (Department of Physics and Astronomy) and Finnish Centre of Astronomy with ESO (FINCA), University of Turku, FI-20014 Turku, Finland; Astronomy Division, University of Oulu, FI-90014 Oulu, Finland} \\
$^{20}$ {Departament de F\'isica, and CERES-IEEC, Universitat Aut\`onoma de Barcelona, E-08193 Bellaterra, Spain} \\
$^{21}$ {Universitat de Barcelona, ICCUB, IEEC-UB, E-08028 Barcelona, Spain} \\
$^{22}$ {ICRANet-Armenia at NAS RA, 0019 Yerevan, Armenia} \\
$^{23}$ {Japanese MAGIC Consortium: ICRR, The University of Tokyo, 277-8582 Chiba, Japan; Department of Physics, Kyoto University, 606-8502 Kyoto, Japan; Tokai University, 259-1292 Kanagawa, Japan; RIKEN, 351-0198 Saitama, Japan} \\
$^{24}$ {Inst. for Nucl. Research and Nucl. Energy, Bulgarian Academy of Sciences, BG-1784 Sofia, Bulgaria} \\
$^{25}$ {Humboldt University of Berlin, Institut f\"ur Physik D-12489 Berlin Germany}\\
$^{26}$ {also at Dipartimento di Fisica, Universit\`a di Trieste, I-34127 Trieste, Italy}\\
$^{27}$ {also at Port d'Informaci\'o Cient\'ifica (PIC) E-08193 Bellaterra (Barcelona) Spain} \\
$^{28}$ {also at INAF-Trieste and Dept. of Physics \& Astronomy, University of Bologna}\\

\label{lastpage}
\end{document}